\documentclass[twocolumn,showpacs,amsmath,amssymb,prl,superscriptaddress]{revtex4}
\usepackage{graphicx}
\usepackage{dcolumn}
\usepackage{bm}
\usepackage{dcolumn}

\begin{document}

\preprint{Draft v4.5}

\title{Parity-violating Electron Deuteron Scattering \\
and the Proton's Neutral Weak Axial Vector Form Factor}

\affiliation {W.K.Kellogg Radiation Laboratory, California Institute
  of Technology Pasadena, CA 91125} 
\affiliation{Department of Physics, University of Illinois at
  Urbana-Champaign, Urbana, IL 61801}
\affiliation{Department of Physics, University of Maryland, College
  Park, MD 20742}
\affiliation{Bates Linear Accelerator Center, Laboratory for Nuclear
  Science, Massachusetts Institute of Technology, Middleton,
  MA 01949}
\affiliation{Department of Physics, Virginia Polytechnic Institute and
  State University, Blacksburg, VA 24061}
\affiliation{Department of Physics and Astronomy, University of
  Kentucky, Lexington, KY 40506}
\affiliation{Physics Division, Argonne National Laboratory, Argonne,
  IL 60439}
\affiliation{Department of Physics, Louisiana Tech University, Ruston,
  LA 71272}
\affiliation{Department of Physics, College of William and Mary,
  Williamsburg, VA 23187}
\affiliation{Laboratory for Nuclear Science, Massachusetts Institute
  of Technology, Cambridge, MA 02139}
\affiliation{LPSC-Grenoble, IN2P3/CNRS-UJF, 38026, Grenoble, France}
\affiliation{Thomas Jefferson National Laboratory, Newport News, VA 23606\\
 and Physics Department, Old Dominion University, Norfolk, VA 23529}

\author{T.~M.~Ito}
\email{tito@krl.caltech.edu}
\affiliation {W.K.Kellogg Radiation Laboratory, California Institute
  of Technology Pasadena, CA 91125} 

\author{T.~Averett} 
\affiliation{Department of Physics, College of William and Mary,
  Williamsburg, VA 23187}

\author{D.~Barkhuff}
\affiliation{Bates Linear Accelerator Center, Laboratory for Nuclear
  Science, Massachusetts Institute of Technology, Middleton,
  MA 01949} 

\author{G.~Batigne}
\affiliation{LPSC-Grenoble, IN2P3/CNRS-UJF, 38026, Grenoble, France}

\author{D.~H.~Beck}
\affiliation{Department of Physics, University of Illinois at
  Urbana-Champaign, Urbana, IL 61801} 

\author{E.~J.~Beise}
\affiliation{Department of Physics, University of Maryland, College
  Park, MD 20742} 

\author{A.~Blake}
\affiliation {W.K.Kellogg Radiation Laboratory, California Institute
  of Technology Pasadena, CA 91125} 

\author{H.~Breuer}
\affiliation{Department of Physics, University of Maryland, College
  Park, MD 20742} 

\author{R.~Carr}
\affiliation {W.K.Kellogg Radiation Laboratory, California Institute
  of Technology Pasadena, CA 91125} 

\author{B.~Clasie}
\affiliation{Laboratory for Nuclear Science, Massachusetts Institute
  of Technology, Cambridge, MA 02139}

\author{S.~Covrig}
\affiliation {W.K.Kellogg Radiation Laboratory, California Institute
  of Technology Pasadena, CA 91125} 

\author{A.~Danagoulian}
\affiliation{Department of Physics, University of Illinois at
  Urbana-Champaign, Urbana, IL 61801} 

\author{G.~Dodson}
\affiliation{Bates Linear Accelerator Center, Laboratory for Nuclear
  Science, Massachusetts Institute of Technology, Middleton,
  MA 01949} 

\author{K.~Dow}
\affiliation{Bates Linear Accelerator Center, Laboratory for Nuclear
  Science, Massachusetts Institute of Technology, Middleton,
  MA 01949} 

\author{D.~Dutta}
\affiliation{Laboratory for Nuclear Science, Massachusetts Institute
  of Technology, Cambridge, MA 02139}

\author{M.~Farkhondeh}
\affiliation{Bates Linear Accelerator Center, Laboratory for Nuclear
  Science, Massachusetts Institute of Technology, Middleton,
  MA 01949} 

\author{B.~W.~Filippone}
\affiliation {W.K.Kellogg Radiation Laboratory, California Institute
  of Technology Pasadena, CA 91125} 

\author{W.~Franklin}
\affiliation{Bates Linear Accelerator Center, Laboratory for Nuclear
  Science, Massachusetts Institute of Technology, Middleton,
  MA 01949} 

\author{C.~Furget}
\affiliation{LPSC-Grenoble, IN2P3/CNRS-UJF, 38026, Grenoble, France}

\author{H.~Gao}
\affiliation{Laboratory for Nuclear Science, Massachusetts Institute
  of Technology, Cambridge, MA 02139}

\author{J.~Gao}
\affiliation {W.K.Kellogg Radiation Laboratory, California Institute
  of Technology Pasadena, CA 91125} 

\author{K.~Gustafsson}
\affiliation {W.K.Kellogg Radiation Laboratory, California Institute
  of Technology Pasadena, CA 91125} 

\author{L.~Hannelius}
\affiliation {W.K.Kellogg Radiation Laboratory, California Institute
  of Technology Pasadena, CA 91125} 

\author{R.~Hasty}
\affiliation{Department of Physics, University of Illinois at
  Urbana-Champaign, Urbana, IL 61801} 

\author{A.~M.~Hawthorne-Allen}
\affiliation{Department of Physics, Virginia Polytechnic Institute and
  State University, Blacksburg,  VA 24061}

\author{M.~C.~Herda}
\affiliation{Department of Physics, University of Maryland, College
  Park, MD 20742} 

\author{C.~E.~Jones}
\affiliation {W.K.Kellogg Radiation Laboratory, California Institute
  of Technology Pasadena, CA 91125} 

\author{P.~King}
\affiliation{Department of Physics, University of Maryland, College
  Park, MD 20742} 

\author{W.~Korsch}
\affiliation{Department of Physics and Astronomy, University of
  Kentucky, Lexington, KY 40506}

\author{S.~Kowalski}
\affiliation{Laboratory for Nuclear Science, Massachusetts Institute
  of Technology, Cambridge, MA 02139}

\author{S.~Kox}
\affiliation{LPSC-Grenoble, IN2P3/CNRS-UJF, 38026, Grenoble, France}

\author{K.~Kramer}
\affiliation{Department of Physics, College of William and Mary,
  Williamsburg, VA 23187}

\author{P.~Lee}
\affiliation {W.K.Kellogg Radiation Laboratory, California Institute
  of Technology Pasadena, CA 91125} 

\author{J.~Liu}
\affiliation{Department of Physics, University of Maryland, College
  Park, MD 20742} 

\author{J.~W.~Martin}
\affiliation {W.K.Kellogg Radiation Laboratory, California Institute
  of Technology Pasadena, CA 91125} 

\author{R.~D.~McKeown}
\affiliation {W.K.Kellogg Radiation Laboratory, California Institute
  of Technology Pasadena, CA 91125} 

\author{B.~Mueller}
\affiliation{Physics Division, Argonne National Laboratory, Argonne,
  IL 60439}

\author{M.~L.~Pitt}
\affiliation{Department of Physics, Virginia Polytechnic Institute and
  State University, Blacksburg,  VA 24061}

\author{B.~Plaster}
\affiliation{Laboratory for Nuclear Science, Massachusetts Institute
  of Technology, Cambridge, MA 02139}

\author{G.~Qu\'{e}m\'{e}ner}
\affiliation{LPSC-Grenoble, IN2P3/CNRS-UJF, 38026, Grenoble, France}

\author{J.-S.~R\'{e}al}
\affiliation{LPSC-Grenoble, IN2P3/CNRS-UJF, 38026, Grenoble, France}

\author{J.~Ritter}
\affiliation {W.K.Kellogg Radiation Laboratory, California Institute
  of Technology Pasadena, CA 91125} 
\affiliation{Department of Physics, Virginia Polytechnic Institute and
  State University, Blacksburg,  VA 24061}

\author{J.~Roche}
\affiliation{Department of Physics, College of William and Mary,
  Williamsburg, VA 23187}

\author{V.~Savu}
\affiliation {W.K.Kellogg Radiation Laboratory, California Institute
  of Technology Pasadena, CA 91125} 

\author{R.~Schiavilla}
\affiliation{Thomas Jefferson National Laboratory, Newport News, VA 23606\\
 and Physics Department, Old Dominion University, Norfolk, VA 23529}

\author{J.~Seely}
\affiliation{Laboratory for Nuclear Science, Massachusetts Institute
  of Technology, Cambridge, MA 02139}

\author{D.~Spayde}
\affiliation{Department of Physics, University of Illinois at
  Urbana-Champaign, Urbana, IL 61801} 

\author{R.~Suleiman}
\affiliation{Laboratory for Nuclear Science, Massachusetts Institute
  of Technology, Cambridge, MA 02139}

\author{S.~Taylor}
\affiliation{Laboratory for Nuclear Science, Massachusetts Institute
  of Technology, Cambridge, MA 02139}

\author{R.~Tieulent}
\affiliation{LPSC-Grenoble, IN2P3/CNRS-UJF, 38026, Grenoble, France}

\author{B.~Tipton}
\affiliation {W.K.Kellogg Radiation Laboratory, California Institute
  of Technology Pasadena, CA 91125} 

\author{E.~Tsentalovich}
\affiliation{Bates Linear Accelerator Center, Laboratory for Nuclear
  Science, Massachusetts Institute of Technology, Middleton,
  MA 01949} 

\author{S.~P.~Wells}
\affiliation{Department of Physics, Louisiana Tech University, Ruston,
  LA 71272}

\author{B.~Yang}
\affiliation{Bates Linear Accelerator Center, Laboratory for Nuclear
  Science, Massachusetts Institute of Technology, Middleton,
  MA 01949} 

\author{J.~Yuan}
\affiliation {W.K.Kellogg Radiation Laboratory, California Institute
  of Technology Pasadena, CA 91125} 

\author{J.~Yun}
\affiliation{Department of Physics, Virginia Polytechnic Institute and
  State University, Blacksburg,  VA 24061}

\author{T.~Zwart}
\affiliation{Bates Linear Accelerator Center, Laboratory for Nuclear
  Science, Massachusetts Institute of Technology, Middleton,
  MA 01949} 

\collaboration{The SAMPLE Collaboration}
\noaffiliation

\date{\today}

\begin{abstract}
We report on a new measurement of the parity-violating asymmetry in
quasielastic electron scattering from the deuteron at backward angles
at $Q^2= 0.038$~(GeV/{\it c})$^2$. This quantity provides a determination
of the neutral weak axial vector form factor of the nucleon, which can
potentially receive large electroweak corrections.  The measured
asymmetry $A=-3.51\pm 0.57\;({\rm stat}) \pm 0.58\;({\rm sys}) $~ppm
is consistent with theoretical predictions. We also report on updated
results of the previous experiment at $Q^2 =0.091$~(GeV/{\it c})$^2$,
which are also consistent with theoretical predictions.
\end{abstract}

\maketitle

Parity-violating electron scattering provides a unique probe of the
electroweak structure of the nucleon. It has been well established
that elastic scattering studies yield new and interesting information
on the strange vector matrix elements~\cite{KAP88,MCK89-BEC89}. This
is the basis for a substantial program of experiments at modern
electron accelerator facilities, beginning with the SAMPLE
experiment~\cite{MUE97-SPA00,HAS00} at MIT-Bates.

The primary goal of SAMPLE is to determine the proton's strange
magnetic form factor $G_M^s$ through parity-violating electron
scattering from the proton at backward angles.  However, the
parity-violating asymmetry -- the asymmetry in the scattering cross
section with respect to the helicity of the incident electron -- is
not only sensitive to $G_M^s$, but is also sensitive to the proton's
neutral weak axial form factor. As pointed out in Ref.~\cite{MUS90},
the neutral weak axial form factor as measured in electron scattering,
$G_A^e$, can potentially receive large electroweak corrections that
are absent in neutrino scattering. These corrections include the
anapole moment, which is identified as the effective parity-violating
coupling of a photon to the nucleon. Determining $G_A^e$ is important
not only for a reliable extraction of $G_M^s$, but also because of its
sensitivity to the hadronic effects on the electroweak radiative
corrections. The adequate understanding of such effects is essential
to proper interpretation of other precision electroweak measurements
such as neutron and nuclear $\beta$ decay~\cite{MCK03-RAM03}.
Parity-violating quasielastic electron-deuteron scattering at backward
angles is predominantly sensitive to $G_A^e$ and thus can be used to
determine $G_A^e$~\cite{BEI91}.

The SAMPLE collaboration previously performed an experiment on a
deuterium target (SAMPLE II) as well as on a hydrogen target (SAMPLE
I) at 200~MeV [$Q^2\sim~0.1$~(GeV/{\it c})$^2$]. Combining the results
from these two experiments allows separate determination of $G_M^s$
and $G_A^e$. Our data~\cite{HAS00} indicated that, while the overall
contribution from strange quarks to the proton's magnetic form factor
is small, the size of the electroweak radiative corrections to the
axial form factor is significantly larger than anticipated from
theory~\cite{ZHU00}.

These results stimulated considerable interest among theorists. Many
different processes and effects were studied for their potential
contributions to the axial form factor or to the parity-violating
asymmetry in electron-deuteron scattering. These include the anapole
moment~\cite{ZHU00,MAE00a-MAE00b,RIS00}, nuclear effects including
two-body currents~\cite{DIA01}, and the parity-violating hadronic
interaction~\cite{SCH03a,LIU03}. None of the effects studied here were
significant enough to explain the discrepancy.

In order to experimentally confirm these results, we performed a third
SAMPLE experiment (SAMPLE III), with a deuterium target at a lower
beam energy of 125~MeV [$Q^2=0.038$~(GeV/{\it c})$^2$]. As was the
case for SAMPLE II, the dominant scattering process is quasielastic
scattering, and the asymmetry is predominantly sensitive to $G_A^e$.
Since the parity-violating asymmetry in the cross section is
proportional to $Q^2$ to first order, the expected asymmetry was
roughly 3 times smaller than that for 200~MeV. The cross section,
however, is larger by a factor of 2 with the same background level,
resulting in an experiment sensitive to the same physics with roughly
the same sensitivity but with very different systematics.

The experiment was carried out at the MIT Bates Linear Accelerator
Center. The experimental method and apparatus were identical to SAMPLE
II, except for the incident beam energy.  A 125 MeV longitudinally
polarized electron beam was incident on a 40~cm long liquid deuterium
target, and electrons scattered at backward angles were detected by an
air \v Cerenkov detector covering angles between 130$^{\circ}$ and
170$^{\circ}$ (solid angle $\sim 1.5$~sr).  The detector consists of
the radiator air volume and 10 detector elements, each with an
ellipsoidal mirror to focus \v Cerenkov light onto a corresponding
8-inch photomultiplier tube (PMT).

A remotely controlled light shutter was used to cover each PMT for the
background measurements. About 10\% of the data were taken with the
shutters closed. In addition, in order to further study the
background, an additional measurement of the shutter closed asymmetry
was made with a plate of plastic scintillator placed in front of each
PMT to enhance the statistics. Also, the non-\v Cerenkov sources of
light in the detector signal, mostly due to scintillation light in the
air, were studied by covering the mirrors. The background level was
the same as SAMPLE II, consistent with bremsstrahlung radiation in the
target being the dominant origin of all the different background
components.

The incident electron beam was pulsed at 600~Hz, and the average beam
current was 40~$\mu$A. The polarized electron beam was generated by
directing a circularly polarized laser beam onto a GaAs crystal.  
The helicity of the beam was pseudorandomly chosen for each pulse.  The
helicity of the beam with respect to the electronic signal was
manually reversed every 2-3 days by inserting and removing a halfwave
plate in the laser beam path to check for and reduce possible
systematic effects. (These two configurations are called ``{\sc in}''
and ``{\sc out}''.)  The polarization of the electron beam was
measured daily with a transmission polarimeter, and occasionally with
the M\o ller polarimeter on the beam line.

Various beam parameters, including the intensity, energy, position,
and angle, were monitored continuously for helicity correlated
differences.  Four Lucite \v Cerenkov counters (luminosity monitors)
downstream of the target at the forward angles ($\sim 12^{\circ}$)
detected low $Q^2$ scattering which has negligible parity-violating
asymmetry, thus serving as monitors for false asymmetries.

\begin{figure}
\resizebox{19pc}{!}{\includegraphics{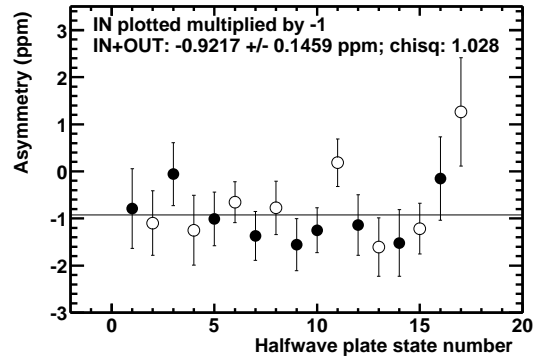}}
\caption{Results for the detector yield asymmetry (average of 10
detectors). Each data point represents the average between two
halfwave plate state changes. The {\sc out} data are plotted as filled
circles and the {\sc in} data as open circles. The sign of the {\sc
in} data is reversed to give the correct sign.\label{fig:superfraser}
}
\end{figure}

As in the past, the yield for each detector for each beam pulse
(integrated over the duration of the pulse) was normalized to the beam
charge measured in front of the target, and then was corrected for the
beam position, angle, and energy.  A linear regression technique was
used to determine the dependence of the detector yields on each
parameter.  Then, the asymmetry was computed for the appropriate pulse
pairs. In addition, the normalized detector yields were also corrected
for transmission of the beam (defined as the ratio of the beam
intensities at the target and at the end of the accelerator) to
account for an observed beam intensity asymmetry caused by
differential scraping at an energy-defining slit. The results of this
analysis are shown in Fig.~\ref{fig:superfraser}, where the detector
asymmetry is plotted as a function of time. 

The asymmetry was further corrected for the beam polarization, the
background dilution, and electromagnetic radiative effects (effects
due to the bremsstrahlung radiation of the incident and scattered
electrons) to obtain the physics asymmetry. The average beam
polarization during the experiment was $P_e = 38.9 \pm 1.6$\%.  The
background dilution factor, determined for each detector from the
ratio between the shutter open and closed detector yields and the
mirror covered studies, was typically $1.4\sim 1.7$ with a relative
uncertainty of 4.5\%. Electromagnetic radiative effects were evaluated
using a spin-dependent modification to Ref.~\cite{MO69} within the
context of a {\sc geant}~\cite{GEANT} simulation of the detector
geometry. In the simulation, scattered electron events were generated
uniformly in energy, angle, and along the length of the 40~cm
target. The scattered electron kinematics were selected after
accounting for energy loss in the target. Each event was weighted
according to the scattering cross section and the detector efficiency,
and was assigned an asymmetry according to its kinematics. The
correction factor for electromagnetic radiative effects was evaluated
for each detector by comparing the (weighted) asymmetry with and
without the radiative effects included in the simulation, and was
typically 1.09 with a relative uncertainty of 3\%.

The systematic error in the corrections procedure was estimated by
comparing results from two different methods that are mathematically
equivalent for normally distributed infinite data: one computes the
dependence of the detector signal on the beam parameters for
normalized yields, and the other for asymmetries. We assign a relative
systematic error of 11.2\% for {\sc out} and 2.1\% for {\sc in}, the
larger error for {\sc out} naturally reflecting the larger correction
due to the larger beam intensity asymmetry.

Additional uncertainties were assigned to the resulting physics
asymmetry to account for two systematic effects observed during the
experiment.  The first is the residual asymmetry in the luminosity
monitors. Some of the luminosity monitors showed non-zero asymmetries
even after the corrections procedure was applied, potentially
indicating the existence of a helicity correlated difference in some
unmeasured beam parameter(s) that caused false asymmetries in the
luminosity monitor signal.  The size of the false asymmetry that this
effect could cause in the \v Cerenkov detector signal was estimated
from the observed luminosity monitor asymmetries and the correlation
between the \v Cerenkov detector asymmetry and the luminosity monitor
asymmetry, and was assigned as the systematic error. Relative
systematic errors of 20.0\% and 19.2\% were assigned for the {\sc out}
and {\sc in} data, respectively, and the errors were treated as
uncorrelated when combining the two data sets.

The second is that, although the measured shutter closed asymmetry for
all 10 detectors combined was consistent with zero, the individual
detectors showed a non-zero shutter closed asymmetry. The
detector-by-detector distribution showed a definite pattern dependent
on the azimuthal angle, indicating that this asymmetry is of
parity-conserving nature, and hence cancels out when averaged over all
10 detectors that are symmetrically arranged azimuthally.  The shutter
closed asymmetry was estimated from the ``high-statistics'' shutter
closed data taken with plastic scintillator and was subtracted from
the shutter open asymmetry for each detector. The value of the final
asymmetry is very insensitive to this procedure because of the
symmetry of the detector arrangement, and the associated systematic
error was estimated to be 5\%.

The resulting physics asymmetry is
\begin{equation}
\label{eq:aphys125}
A(Q^2=0.038) = -3.51\pm 0.57\pm 0.58\;\;{\rm ppm},
\end{equation}
where the first uncertainty is statistical and the second is the
estimated systematic error as summarized in
Table~\ref{tab:errorbreakdown}.
\begin{table}
\caption{Summary of relative uncertainties on the measured
asymmetry.\label{tab:errorbreakdown}}
\begin{ruledtabular}
\begin{tabular}{lc}
\multicolumn{1}{c}{Source} & $\delta A/A$~(\%) \\ \hline
Dilution factor & {\  5}\\ 
Background asymmetry subtraction & {\  5}\\ 
Luminosity monitor asymmetry & 14\\ 
Corrections procedure & {\  5}\\ \hline
Total systematic (added in quadrature) & 17 \\
\end{tabular}
\end{ruledtabular}
\end{table}

Since the SAMPLE detector does not have energy resolution for the
scattered electrons, the measured asymmetry contains contributions not
only from quasielastic scattering but also from elastic scattering and
threshold breakup. In order to construct the theoretical expression of
the asymmetry as a function of the quantities of interest, {\it i.e.},
$G_M^s$ and $G_A^e$ of the nucleon, we did the following. First we
performed a full nuclear calculation according to Ref.~\cite{SCH03a}
to obtain the parity-conserving and parity-violating response
functions for the total inelastic processes (quasielastic scattering
and threshold breakup) for selected kinematics.  The dependence on
$G_M^s$ and $G_A^e$ was explicitly kept track of in the
calculation. Electroweak radiative corrections were included. In
particular, the isoscalar axial radiative correction was taken to be
$R_A^0=0.03\pm 0.05$ from Ref.~\cite{ZHU00}. We use
$\sin^2\theta_W=0.23113(15)$~\cite{PDG}.

The parity-violating asymmetry was computed on an event-by-event basis
in the {\sc geant} simulation, and separately for the elastic (from
Ref.~\cite{POL90}) and inelastic (using the above obtained response
functions) processes. The resulting
asymmetry distributions represented an average over the detector
acceptance and incident electron energies. The physics asymmetry was
then computed as a combined average of the elastic and inelastic
distributions weighted by the appropriate cross sections. The
resulting theoretical asymmetry is
\begin{equation}
A(Q^2=0.038) = -2.14 + 0.27 G_M^s + 0.76 G_A^{e\;(T=1)},
\end{equation}
where the asymmetry is in parts per million and the form factor is in
nuclear magnetons (n.m.). In this expression, we retain explicitly the
isovector ($T=1$) component of $G_A^e$. The small isoscalar component
is absorbed into the first term. The dependence on the nuclear model
is small.

The radiative corrections and theoretical asymmetry for the SAMPLE II
data were also re-evaluated with the {\sc geant} simulation. In
addition, background dilution factors coming from pion photoproduction
were re-examined in light of recently published
data~\cite{BER98}. Such processes contribute to the detector yield
through their decay products, but have negligible parity-violating
asymmetry~\cite{CHE01}. The largest contribution is from coherent
$\pi^0$ photoproduction on the deuteron, which had been neglected in
Ref.~\cite{HAS00}, but was found in Ref.~\cite{BER98} to be
significantly enhanced relative to the corresponding incoherent
process. Including this effect increased the background dilution
factor by 9\%. The re-evaluated electromagnetic radiative corrections
resulted in an additional 2\% increase in the background dilution
factor. Finally, improved determination of the scintillation component
of the detector signal resulted in another 2\% increase. Thus, the
final physics asymmetry increased by 13\% in magnitude compared with
our previously published results~\cite{HAS00}, giving
\begin{equation}
A(Q^2=0.091) = -7.77 \pm 0.73 \pm 0.62\;\;{\rm ppm},
\end{equation}
where the first error is statistical and the second is the estimated
systematic error. (Note that this asymmetry value contains the
contribution from the non-quasielastic processes, which was estimated
to be $\sim 1.5$\% and removed in Ref.~\cite{HAS00}.)  The
re-evaluated theoretical value for the asymmetry, using the nuclear
calculation as described above, has resulted in an expected value that
is 2\% smaller:
\begin{equation}
A(Q^2=0.091) = -7.06 + 0.77 G_M^s + 1.66 G_A^{e\;(T=1)}.
\end{equation}

\begin{figure}
\resizebox{19pc}{!}{\includegraphics{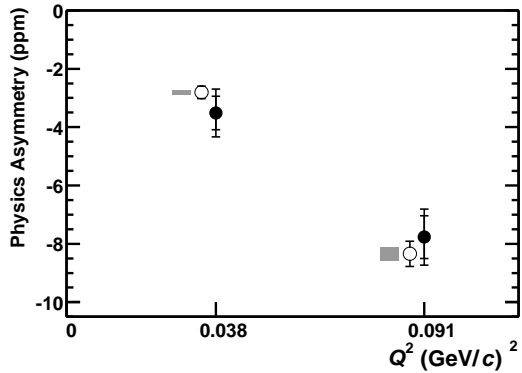}}
\caption{The physics asymmetries measured in SAMPLE II (updated
results) and SAMPLE III are plotted as a function of $Q^2$ (solid
circles). Also plotted (with offset $Q^2$ for visibility) are the
theoretical predictions with the value of $G_A^e$ taken from
Ref.~\protect{\cite{ZHU00}}, and $G_M^s=0.15$~n.m. (open circles). The
height of the gray rectangles represents the change in the physics
asymmetry corresponding to a 0.6~n.m. change in
$G_M^s$. \label{fig:AphysvsQ2} }
\end{figure}
In Fig.~\ref{fig:AphysvsQ2}, the physics asymmetries measured in
SAMPLE II (updated results) and SAMPLE III are plotted as a function
of $Q^2$. Also plotted are the theoretical predictions with the value
of $G_A^e$ taken from Ref.~\cite{ZHU00} [$G_A^e(Q^2=0.038)=-0.91 \pm
0.28$ and $G_A^e(Q^2=0.091)=-0.84 \pm 0.26$], and
$G_M^s=0.15$~n.m. The dependence of the theoretical values on $G_M^s$
is small.

The results from SAMPLE III (125 MeV deuterium run) and the updated
results from SAMPLE II (200 MeV deuterium run) both agree with the
theoretical prediction on the electroweak radiative correction on the
neutral weak axial form factor of the nucleon by Zhu {\it et
al.}~\cite{ZHU00}. In addition to these two experimental results,
various theoretical efforts also support the theoretical prediction by
Zhu {\it et al.}  The confirmation on the theoretical value of
$G_A^e$ not only allows us to extract $G_M^s$ reliably from the data
from SAMPLE I (200 MeV hydrogen run), but also is important for
interpreting results from future parity-violating electron scattering
experiments at JLab and Mainz.

We gratefully acknowledge the skillful efforts of the staff of the
MIT-Bates laboratory to support the experiment and provide a high
quality beam.  We thank M.~J.~Ramsey-Musolf for insightful
discussions.  This work was supported by NSF, DOE, Jeffress Memorial
Trust, and CNRS of France.

\end{document}